# A Fundamentally New Perspective on the Origin and Evolution of Life

Shi V. Liu

Eagle Institute of Molecular Medicine
Apex, NC 27502, USA

SVL@logibio.com

**Motto**

*Criticism and attempted "falsification" are essential parts of science.*
- Philip J. Darlington, Jr. (1904-1983)

## HIGHLIGHT

A fundamentally different perspective on the origin and evolution of life was provided over 17 years ago and should be seriously considered today by all researchers in the field of evolution study.

## ABSTRACT

Darwin's hypothesis that all extant life forms are descendants of a last common ancestor cell and diversification of life forms results from gradual mutation plus natural selection represents a mainstream view that has influenced biology and even society for over a century. However, this Darwinian view on life is contradicted by many observations and lacks a plausible physico-chemical explanation. Strong evidence suggests that the common ancestor cell hypothesis is the most fundamental flaw of Darwinism. By contrast, a totally different perspective on origin and evolution of life claims that cellular life forms were descendants of already diversified acellular life forms. Independently originated life forms evolve largely in some parallel ways even though they also interact with each other. Some evolutionary "gaps" naturally exist among evolutionary lines. Similarity may not be the only result of phylogenetic inheritance but may be a result of a convergent mechanism of origin and evolution. Evolution is not a random process but follows some basic physico-chemical principles as a result of the interplay of both energy and entropy on matter.

Next year will be the 150[th] anniversary of the publication of Darwin's famous book "*On the Origin of Species by Means of natural Selection or the Preservation of Favored Races in the Struggle for Life*" (1) and a celebration for the 200[th] birthday of a most influential biological scientist of all time (2, 3).  There is no doubt that Darwin's view has contributed greatly to the winning of a strong voice for science in explaining the most interesting but also the most perplexing question that humanity has confronted with – when, where, and how life originated and evolved.  However, is Darwinian view really a truthful reflection of the natural history of life?  Are there any other perspectives that worth of consideration and may ultimately be proven more compelling than Darwinism in explaining the origin and evolution of life?

　　　Presented here is an abstractive summarization of various mistakes contained in Darwinism which include not just Darwin's own original works but also many add-on contributions from his followers.  Based on this collective criticism on Darwinism, a fundamentally different perspective on the origin and evolution of life is introduced.  This different perspective was introduced into public in 1991 and formally published in 2006 (4).  However, considering the lack of recognition for that publication and the emergence of more evidence, it is necessary to represent this perspective in a more noticeable way with some updated information.

**Fundamental flaws of Darwinism**

Due to the dominance of Darwinism in evolution study, it is difficult to directly express a different perspective without proving a minimal necessity of offering that perspective.  That minimal necessity may come with a realization that Darwinism, no matter how popular it is, is still far from a complete understanding of life history.

　Over the past time there have been many criticisms on Darwinism (5-7).  It is beyond the scope of this communication to summarize all of these criticisms.  However, the following list may show some major shortcoming and inconsistency in the Darwinian view of the life history.

*1. Gradual evolution was not backed by fossil records.*

　The history of most fossil species includes two features particularly inconsistent with gradualism as conceived by Darwinian evolutionists (8-13): a stasis during which time most species exhibit no directional change and a sudden appearance of different species in many local areas as "fully formed".  Evolutionary process also appears to have different phases, some periods having considerable evolutionary activity and some periods being comparatively quiescent.

*2. Evolutionary progress is not a random process.*

　Evolution has been a one direction irreversible process as shown by the fossil records which is inconsistent with prediction made by randomness (5, 14, 15).  Order prevails in the formation of protocell as it is shown that amino acids turned to be "self-sequencing" and DNA and RNA are built of a phosphate and sugar backbone (16).  Thus when cells come into existence, they come in a way different from that foreseen by randomness.

　As stated by Darlington, all processes of directional change (which include evolution of life) are programmed to some extent by the composition and limits of the sets concerned, and by environmental factors, even though complex evolutionary processes are at the same time partly open-ended (17).

*3. Gene mutation is not the main overall mechanism of evolution.*

　Evolution of biochemistry preceded the innovation of genetic system.  Form and function arise from the physico-chemical and mineral imprints of the previous levels of evolution.  Genes and chromosomes influence form and function of life but this is not tantamount to them being the originators of these processes.  Their intervention is at a secondary level, deciding only what variant of form and function will become fixed.  Mutation alone cannot explain the increased complexity as reflected in the increase of genome size.  Thus, it is impossible to understand evolution of coordinated complexities purely in terms of piecemeal accumulation of random mutations.  As a matter of fact, attempts in dating evolutionary process using the molecular clock driven by gene mutation have generated more inconsistence than consistence (11, 18, 19).  Alarm was even raised: molecular clocks run out of time (20).

*4. Natural selection is a vague term and should be better replaced with a concrete physico-chemical mechanism.*

　Natural selection is currently defined as the process by which favorable heritable traits become more common in successive generations of a population of reproducing organisms, and unfavorable heritable traits become less common, due to differential reproduction of genotypes.  In the original words by Darwin, natural selection is defined as "preservation of favorable variations and the rejection of injurious variations" (1).

This definition describes an outcome of a process rather than a mechanism for a process. As a matter of fact, the original concept of natural selection was developed in the absence of a valid theory of inheritance because nothing was known about genetics at the Darwinian time. As pointed out by Darlington, "almost all writers on evolution use natural selection, but hardly anyone tells us what it is or how it works"(21).

To be meaningful for evolutionary study, it is essential to replace a vague term "natural selection" with a concrete physico-chemical process so its actual interaction with a living organism can be measured in well-defined units (22, 23).

Darlington has argued that natural selection is best defined as *differential elimination.* He pointed out that stressing differential reproduction alone is a distortion. This is because *differential elimination* happens at all levels of evolutionary processes such as differential survival of reproduced individuals (17).

*5. The notion of "common origin" from a last common ancestor cell may be the most erroneous guess.*

Theoretically it is unbelievable that only one ancestor cell formed in a life-forming environment that seemed to prevail in an enormous space. It might also be extremely difficult for a single cell to proliferate fast enough to account for the rapid appearance of different life forms and the wide distribution of living organisms on Earth. It has not been shown why the common ancestor cell should diversify into many other forms if they lived in the same environment or why a same global change of environment could have a differential selection so that many different progenies produced and stabilized.

It should be emphasized that "similarity" in life forms can also be resulted from truly independent events taking place according to a common mechanism. The current phylogenetic studies have not rigorously rule out this possibility in their tree constructions. And there is no reason to deny that there may be more than one precursors in pre-cellular phases such as the surface metabolism, the RNA world, the RNP world and the DNA world that would independently "cellularized" (24).

*6. Current mainstream evolutionary theories separated the evolution of Earth from the evolution of life.*

This separation occurs because biological evolution is based on the mutation of genes which does not belong to inorganic non-living world. This has created a mysterious cloud around the biological subjects and delayed the approaches of deciphering the biological world by the same laws that governed the abiotic world. However, as we look into the deeper level of biology, we actually see more and more connections between the biotic world and the abiotic world (25-28). In the words of Darlington, prebiotic and biotic evolution on the earth's surface is a continuous, energy-requiring process (17).

**A fundamentally different perspective on evolution**

As a result of the above literature review a new perspective on the origin and evolution of life was proposed in 1991. The original paper was entitled "Evolution: an integrated theory - Criticisms on Darwinism" which was a thesis for my graduate course on microbial evolution. It was published only much later in 2006 for an obvious reason (4).

Summarized here are the key points of that "new" but actually 18 years-old perspective on evolution, with some updated information:

*1. Evolution consists of two organically connected sides: the abiotic evolution and the biotic evolution.*

Evolution is a phenomenon inherent to the construction of the universe. It actually starts with the formation of elementary particles at the dawn of the conversion of energy into matter (29). Evolution starts when the universe is born. Evolution cannot be considered solely as a biological process. The elementary particles, the chemical elements and the minerals have each had an autonomous evolution (25, 29). Biotic evolution was anteceded by these three levels of evolution. Evolution of life on Earth is one component of the overall development of the Earth. The biotic and abiotic worlds are organically connected and have mutual influence on each other's development (17, 26, 30-33).

*2. Evolution is governed by some fundamental laws.*

Evolution of both abiotic and biotic worlds is governed by some common fundamental laws (34). Evolution is an entropic phenomenon, a natural consequence of the behavior of physical information arrays in an entropic universe (35). Evolution in general or totality is characterized by increasing entropy as evidenced by increasing complexity and diversity. Thus, evolution is result of the interplay between energy and entropy and follow some basic laws such as the second law of thermodynamics (35, 36).

Speaking of biotic evolution, some people have argued that entropy rules out evolution because the order produced within cells is more than that compensated for by the disorder they create (37). Indeed, energy or order is usually associated with normal life and those molecular guardians that ensure it, but entropy or disorder has been associated with disease (38). Such postulated prevention of or yet resistance to entropy in order to maintain normal life and thus ultimately direct evolution has meanwhile been recognized by several investigators (23).

However, it must be pointed out that whether or not a thermodynamic law can be directly applied is also dependent on the nature of the system and its reference to the surrounding environment. In some circumstances, it even depends on how the law is defined or interpreted. For example, the second law of thermodynamics states the total entropy of any isolated thermodynamic system tends to increase over time. The universe is an isolated system. Thus its total entropy would naturally increase. However, individual life is not an isolated system but is openly connected with its environment. Thus, any consideration of entropy balance and the correctness/applicability of physical laws must take into account of the nature of the system (39).

When the totality of life forms in an isolated universe is considered, we must realize that it is the dispersal of energy (another way to define *entropy*) from initially fewer forms of life to later on more forms of life that captured a global picture of evolution over the geological time. This change in the biotic world is consistent with an increase of entropy if we agree that the emergence of new forms of life is either due to a mutation (a kind of disorder) of one life form or a mixing of different life forms (a kind of entropy mixing which leads to an increase of entropy in the system even if there is not net exchange of heat or work).

Thus, I still firmly believe that "any theory claiming to describe how organisms originate and continue to exist by natural causes must be compatible with the first and second laws of thermodynamics." (40).

*3. Evolutionary process is a network of consecutive loops of current flow reactions.*

Currents of energy flow and mass flow are common to both abiotic and biotic systems. The global current flow can be resolved into a series of current loops originating at source points and terminating at sink points (41). Each source serves to drive cycles of matter around the reaction loops. The beginning of the Earth system is the network in small molecule space. The transition from small molecule space to macromolecular space and then to organism space represents a reflexive autocatalysis network of loops. The global pre-biotic ecological cycles, with Sun being the radiation source and outer space being the infinite isothermal reservoir, provides the functional basis for biogenesis.

*4. Evolution is an ordered process and the pattern realized in the foregoing loops channeled the development of subject in the following loops.*

As the system has aged it has explored Platonic space in a highly directed way (highly nonrandom). The reactions in the foregoing loop canalize the potential of the reactions in the following loops. The central problem of evolution is not the "origin of species" but the origin of form and function. The biotic evolution was preceded by the abiotic evolution (42) and thus it cannot depart from the frame created by the physico-chemical laws and rules. Formation of different life forms is a type of events determined by the physico-chemical identity of corresponding pro-cellular matters (43). The process of evolution is not a random process (14, 29, 43, 44).

Matter takes the forms of life because it has the inherent capacity to do so. The kinds and numbers of genes available to form sets (genotypes) are limited and selection even further eliminate "unfit" combinations (17). Thus, even though the overall evolution process can be considered as a kind of losing order and thus increase of entropy, each individual evolution process still follow some specific mechanisms intrinsically determined by the matters comprising that line of evolution.

*5. Life forms had originated at multiple places and may have arisen multiple times.*

Life forms, even the same form of life, can be formed at multiple locations on Earth as long as the environmental conditions that favored their formation are the same. Different life forms may arise in different geological times (45).

Arguments have been made on what the earliest life form was (46-50). However, any argument on this matter must be discussed in the context of contemporary environmental conditions and how the emergence of the biotic forms alters the abiotic world.

Furthermore, when discussing how living beings emerged from abiotic components (in similar ways at different locations), it is worth mentioning a more recent hypothesis on a peptide origin of life (51). It postulates that secondary to the formation of oligopeptides from amino acids under conducive prebiotic conditions such as those involving (volcanic) carbonyl sulfide as a catalyst (52), further peptide bond formations engender the production of significant amounts of water which along with carbon dioxide could be processed in photosynthetic reactions to generate oxygen and those higher organic matter building blocks that are necessary for the emergence of life (51).

*6. Biotic evolution showed distinctive phases which had its own mechanism and was influenced by the*

*corresponding abiotic evolution.*

The history of the biological evolution may have different phases determined by the abiotic evolution of Earth environment. Each phase may produce specific change(s) of certain components of living system (in other words the variation may be shown in different ways) (53). Thus the reconstruction of its history may not be achieved by using a single universal yardstick (12).

*7. The evolution of different branches of life forms may have different rates and this unbalanced evolution allows the co-existence of so many types of life forms in a contemporary time and common space.*

The evolution potential and the speed of evolution are determined by the distinguished physico-chemical properties of each life form. This independence gave an unbalanced evolution of different life forms. The co-existence of diversified types of the extant and/or extinguished life forms in the contemporary time and common space is a manifestation of such property (9).

The multi-origin and multi-line scenario of life evolution actually adds more dimensions to the thermodynamic operation of evolution. From the very beginning, the increase of entropy of the living world might not just be a result of mutation (disorder) but more likely a result of combination of mutation and mixing (horizontal gene transfer and symbiosis for example). Thus the ruling out of entropy from the evolutionary process simply because life feeds on negative entropy or "negentropy" (54) may be a misunderstanding of the history of life.

Obviously, this new perspective on evolution does not fit into any aspect of the Darwinism because it basically contradicts all the fundamental assumptions withheld by the Darwinian Theory on evolution. However, if we agree with Darlington that "much widely accepted evolutionary and sociobiological theory, presented by the most influential evolutionists, is wrong and therefore dangerous" (21), then everything on evolution should be open to full discussion.

**Increasing new evidence contradicting Darwinism**

Now seventeen years have passed since my initial proposal of an alternative perspective on evolution, was I too naïve in asking some "stupid" questions or did I actually trip over some intelligence on evolution?

Let me take a quick look at some most important developments in recent evolutionary studies.

1. The search for the Last Common Ancestor (LCA) or Last Universal Cellular Ancestor (LUCA) has not lead to any clear answer but more contradictions (46, 55-64). It has even become increasingly hopeless to find any single "root" as more and more evidence now show the deep branches of the Tree of Life (TOL) actually look more like "rings" or "networks" (65-67). What does this mean? There was no common ancestor cell for all the life forms formed later. The cellular life was already a "forest" when we first saw it (4, 68).

2. The application of any single "molecular clock" for calculating the speed of evolution has resulted in repeated timing errors not only against geological time tables but also within biological time tables (69-72). This means that evolution may not be a single line linear process operated under a single mechanism (of gene mutation).

3. Many claimed "missing links" are still missing despite many hard efforts in searching them. Do we have any chance to find these "missing links" or do they really exist or existed-once but then disappeared? Is it actually necessary to find these "missing links" in order to "complete" our reflection of the history of evolution? The answers depend on whether you still believe that there was a LCA or not. When there was no LCA, the "links" are unnecessary because each different pre-cellular ancestor can evolve into different cellular life forms without requiring those intermediate linkages. In other words, there are natural "gaps" between different lines of evolution.

4. Many assumed single line linear evolutions contrast with more and more solid evidence of multi-line coevolution (73-75). Coevolution/co-evolution (76-79) or "collective evolution" (80) has become a prominent theme in evolutionary studies (73, 81-93). But this coevolution may be a natural outcome of the parallel evolution of independently originated multi-lines of life forms. Thus the "pattern pluralism" on the Tree of Life (74, 94) may be a collective perception of related but independently evolved entities rather than diversified entities came from a single origin.

Besides these generalized overviews of the recent progresses in evolutionary studies, I am more encouraged with seeing some solid conclusions contained in some recent publications by the following individuals. For example, Carl Woese has come to a realization that "The universal ancestor is not a discrete entity. It is, rather, a diverse community of cells that survives and evolves as a biological unit" (95). Doolittle also pointed out that "a single tree-like pattern is not the necessary …… Pattern pluralism is an attractive alternative to the quixotic pursuit of a single true TOL."(94). Koonin et al. have concluded that "The last universal common ancestor (LUCA) ……was not free-living but an inorganically housed assemblage of expressed and replicable genetic elements" (96) and discussed the ancient virus world in relation with the evolution of cells (49). Eugene Koonin later also showed that the patterns of life emergence on earth can be better explained by a Biological Big Bang model (97). Patrick Forterre proposed that the transition from RNA to DNA genomes would have stabilized the three canonical versions of proteins involved in translation, whereas the existence of three different founder DNA viruses explains why each domain has its specific DNA replication apparatus (47). Michael Lynch stated that the origins of many aspects of biological diversity, from gene-structural embellishments to novelties at the phenotypic level, have roots in nonadaptive processes (98)

So, evidence conforming to my different perspective on evolution is increasingly incoming. More importantly, many remaining "puzzling observations" and "intriguing conflicts" can be easily resolved with a multi-origin parallel evolution view. For example, the existence for certain pairs of amino acids with an unclear biosynthetic relationship between the precursor and product amino acids and the collocation of Ala between the amino acids Val and Leu belonging to the pyruvate biosynthetic family (73) may not be viewed as a confrontation with the coevolution the occurrence of highly similar genetic modules in diverse organisms needs not to be explained by the assumed highly leaky membranes and the assumed high level of horizontal gene transfer (HGT) (99, 100). This is because some same genetic modules can be sealed into different cells at the major transitions from acellular to cellular life (4, 68).

Thus, considering so many problems with the traditional views on evolution, the old-fashioned view such as the "Darwinism" should be replaced by insightful knowledge (101). I may be wrong in proposing a fundamentally different view than Darwinism. But If I am right, all interested persons should not stubbornly refuse to face my view.

**Conclusions**

The central position of this new perspective on evolution is that cellular life forms might have multiple independent origins that were rooted in different acellular forms. The formation and evolution of life is not random but abides some physico-chemical principles. Each independently originated cellular life form may evolve under separate phylogenetic lineages and may also interact with each other to form more complex higher order life forms. Thus, the images of the life history as reflected by the later reconstruction efforts may appear as "mosaics" and show some pattern pluralisms. The "Tree of Life" (TOL) may serve better as a classification scheme for grouping life forms with similar structure and function rather than as a tracing tool for identifying the phylogenetic relationship and even the so-called the Last Common Ancestor (LCA).

I wish this different perspective on the life history will enhance our understanding about the origin and evolution of life. I sincerely invite all people interested in evolutionary study to voice their opinions because I truly believe debating controversies can enhance creativity (102). Remember, "in the history of science, it is the change in perspective that has usually been the most important" (103).

To celebrate Darwin as a great scientist (3), we should fulfill the dream of a scientist to advance science with continued efforts. This is because, what characterizes science is its search for a logical interpretation of a given event. The agreement between the interpretation and the phenomenon depends on the amount of knowledge available at a given time. Science is not afraid, in the long run, to change completely its outlook and position. Good science provides for a continuing search for new ground and review of old (4).

**Appendixes**

1. **History of this manuscript.**
   This manuscript was evolved from a Comment submitted to *Biology Direct*. Dr. Eugene Koonin has provided many valuable inputs in the transformation of that Comment into a Hypothesis type manuscript. However, this does not necessarily mean that he is in agreement with the views expressed in this manuscript. Request for reviewing this manuscript has been sent to over 13 board members of *Biology Direct* including Drs. Eric Bapteste, Juergen Brosius**,** Ford Doolittle, Patrick Forterre, Peter Gogarten, Simonetta Gribaldo, Gaspar Jekely, Alexey Kondrashov, Doron Lancet, John Logsdon, Bill Martin, Arcady Mushegian and Anthony Poole. However, only one board member sent back a review. Some stated of lack of time to do the review and most other simply ignored the request.
   The manuscript was also sent to Dr. Carl Woese wishing that he can communicate it to *PNAS* or write a review. He declined both requests and suggested a direct submission to *PNAS*. The direct submission was rejected by *PNAS* because "The Board concluded that while interesting, your paper lacked the broad appeal necessary for further consideration".

2. **Review received and agreed to be published.**
   Removed to comply the policy of arXiv.